\documentclass[aip,jcp,amsmath,amssymb,reprint,preprintnumbers]{revtex4-1}
\usepackage{graphicx}
\usepackage{latexsym}
\usepackage{mathrsfs} 
\usepackage{dcolumn}
\usepackage{bm}
\begin{document}


\title{Crowding effect on helix-coil transition:  beyond entropic stabilization}

\author{A. Koutsioubas}
\affiliation{Laboratoire L\'eon Brillouin, CEA/CNRS UMR 12, CEA-Saclay, 91191 Gif-sur-Yvette cedex, France}

\author{D. Lairez}
\email[Corresponding author. E-mail: ]{lairez@cea.fr}
\affiliation{Laboratoire L\'eon Brillouin, CEA/CNRS UMR 12, CEA-Saclay, 91191 Gif-sur-Yvette cedex, France}

\author{S. Combet}
\affiliation{Laboratoire L\'eon Brillouin, CEA/CNRS UMR 12, CEA-Saclay, 91191 Gif-sur-Yvette cedex, France}

\author{G. C. Fadda}
\affiliation{Laboratoire L\'eon Brillouin, CEA/CNRS UMR 12, CEA-Saclay, 91191 Gif-sur-Yvette cedex, France}
\affiliation{Universit\'e Paris 13, UFR SMBH, 93017 Bobigny, France}

\author{S. Longeville}
\affiliation{Laboratoire L\'eon Brillouin, CEA/CNRS UMR 12, CEA-Saclay, 91191 Gif-sur-Yvette cedex, France}

\author{G. Zalczer}
\affiliation{Service de Physique de l'Etat Condens\'{e}, CEA-Saclay, 91191 Gif-sur-Yvette cedex, France.}

\date{\today} 

\begin{abstract}
We report circular dichroism measurements on the helix-coil transition of poly(L-glutamic acid) in solution with polyethylene glycol (PEG) as a crowding agent. The PEG solutions have been characterized by small angle neutron scattering and are well described by the picture of a network of mesh size $\xi$, usual for semi-dilute chains in good solvent. We show that the increase of PEG concentration stabilizes the helices and increases the transition temperature. But more unexpectedly, we also notice that the increase of concentration of crowding agent reduces the mean helix extent at the transition, or in other words reduces its cooperativity. This result cannot be taken into account for by an entropic stabilization mechanism. Comparing the mean length of helices at the transition and the mesh size of the PEG network, our results strongly suggest two regimes: helices shorter or longer than the mesh size.
\end{abstract}
 

\maketitle

Proteins are the functional macromolecules of the cell. Their smooth functioning depends on a determined three dimensional structure usually named as their "folded state". The complete understanding of the way a linear polypeptide chain passes from a disordered and random coil conformation to this folded state, i.e. the protein folding process, is still puzzling and is a major challenge for current biology. In cells, macromolecule crowding and confinement play a central role in the thermodynamics of this folding process\cite{Minton:1983fk}. The main idea lies in an "entropic stabilization mechanism": the excluded volume due the presence of interface (confinement) or other macromolecules (crowding) lowers the conformational entropy of random conformations, which may favor the formation and the stabilization of well organized structures\cite{Zhou:2008, Ziv:2005fk}. In many cases, competition between different possible structures complicates the system and imposes to introduce refinements in order to account for the geometry and size of the confinement space or of the crowding agent\cite{Javidpour:2011fk}. Model and simplified systems are entry points of this complexity. In this context, $\alpha$-helix that is one of the two main structural elements of proteins is particularly interesting, because helix-coil transition can be observed for homo-polypeptides. This allows us to get rid of the polyampholyte feature (presence in the same chain of positively and negatively charged monomers) and of the amphipathic feature (presence in the same chain of hydrophilic and hydrophobic monomers) of proteins. Regarding studies of crowding effect, current trends of \textit{in vitro} experiments are to use an inert crowding agent such as Ficoll, a highly branched polysaccharide that behaves as compact and hard spheres, or polyethylene glycol (PEG) that behaves as a linear polymer in good solvent\cite{Zhou:2008, Stanley2008}.

This paper is concerned with the helix-coil transition of poly(L-glutamic acid) embedded in a semi-dilute solution of long PEG chains, i.e. above their overlap concentration in a regime where they become interwoven to form a network. In this regime\cite{deGennes:1996}, the thermodynamic properties of the solution are independent of the molecular weight and can be described using a single length scale $\xi$ that only depends on the volume fraction of chains. This point is the main difference between previous\cite{Stanley2008} and present works.

We report circular dichroism measurements at different pH and concentrations of PEG as a function of temperature. These measurements give access to the helix fraction $x_h$ of a chain that can be considered as the order parameter of the transition.
In contradiction with recent molecular dynamics simulations\cite{Kudlay:2009kx}, but in agreement with the entropic stabilization mechanism, we show that the helix-coil transition temperature $T^*$ increases with PEG concentration. However, we show that unexpectedly the ratio $T/T^*$ is not a reduced variable for $x_h$, i.e. the curves $x_h(T/T^*)$ obtained for the different PEG concentrations do not superimpose. In the framework of a transition between two states of different energy, this result firstly demonstrates the cooperativity of the helix formation, i.e. the necessity to introduce a coupling parameter in the free energy, and secondly that this feature depends on the PEG concentration. 

The corollary of cooperativity is the existence of correlations in helix distribution. Traditionally, the first and simplest theoretical model introducing this ingredient is the Zimm-Bragg model\cite{zimm:526} that can be mapped into a one dimensional Ising model with an external field\cite{Badasyan2010} and an attractive (ferromagnetic) coupling between adjacent helix turns. By analyzing our data in this framework, we find that the attractive coupling becomes less efficient for increasing concentration of PEG. Our results suggest two regimes depending on the extent of helical domains compared to the mesh size of the PEG network.

\section{Results}
\subsection{Materials and methods}

Poly(L-glutamic acid) (molecular weight $M_w=18$\,kg/mol, polydispersity index 1.02) and polyethylene glycol (molecular weight $M_{\text{PEG}}=20$\,kg/mol, polydispersity index 1.04) where purchased from Sigma Aldrich and were used without any further purification. 

Small angle neutron scattering (SANS) measurements were performed with PACE spectrometer (LLB neutron facility)  at 20$^\circ$C on PEG solutions in heavy water ($^2$H$_2$O) in order to maximize contrast and minimize incoherent scattering. Data reduction was done following ref.\cite{Brulet:2007}.

For circular dichroism (CD) measurements, aqueous solutions (light water $^1$H$_2$O, 10\,mM phosphate buffer, no salt added) containing various concentrations of PEG (up to 30\,wt\%) were vigorously agitated for at least two days in order to ensure complete dilution. The pH of the solutions was adjusted by addition of HCl and NaOH aqueous solutions. Before each measurement a small volume of relatively concentrated solution of poly(L-glutamic acid) was added in order to achieve a final peptide concentration of 10\,$\mu$M. Measurements were conducted using a JASCO J-815 CD spectrometer (CEA/DSV/iBiTec-S/SIMOPRO). Solutions were placed in quartz cuvettes with 1\,mm path length. At the fixed wavelength of 222\,nm, temperature scans in the range of 5-95$^\circ$C (heating and cooling rates equal to 1\,deg/min) were performed using a Peltier device. 

\subsection{SANS characterization of PEG solutions}

\begin{figure}[!htbp]
\centering
\includegraphics[width=1\linewidth]{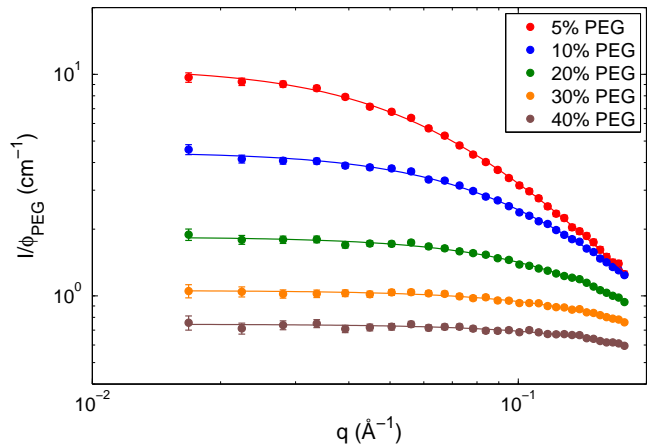}
\caption{Coherent differential scattering cross section per unit volume and unit of PEG volume fraction $I/\phi_{\text{PEG}}$ vs. scattering vector $q$ for PEG solutions at  $T=20^\circ\text{C}$. Lines are best fits following Eq.\ref{eqoz}.}
\label{figspectre}
\end{figure}

\begin{figure}[!htbp]
\centering
\includegraphics[width=1\linewidth]{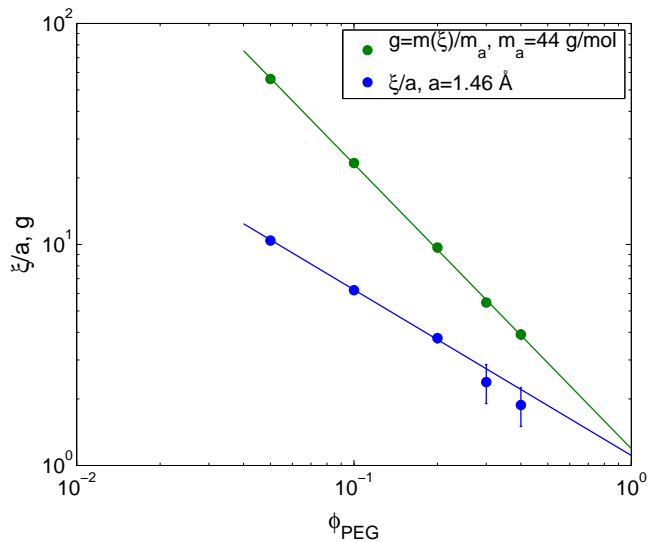}
\caption{Mean short distance, $\xi/a$, and curvilinear distance, $g=m(\xi)/m_a$, between neighboring PEG chains vs. volume fraction $\phi_{\text{PEG}}$, measured at $T=20^\circ\text{C}$, with $m_a$ the molar mass of the monomer and $a$ its length from ref.\cite{Mark:1965fk}. Straight lines are best power law fits corresponding to $\xi/a\propto\phi_{\text{PEG}}^{-0.75\pm 0.03}$ and $g\propto\phi_{\text{PEG}}^{-1.31\pm 0.01}$, respectively.}
\label{figxi}
\end{figure}

Here, the crowding agent is polymer chains in semi-dilute solution. In this concentration regime, the thermodynamic properties of the solution depends on a single length scale, $\xi$, that is independent of the  length of the chains but depends only on their volume fraction\cite{Cotton:1972, Daoud:1975uq}. In semi-dilute solutions the osmotic pressure that writes 
\begin{equation}\label{pi}
\pi=kT/\xi^3
\end{equation}
with $kT$ the thermal energy, is actually measured as being independent of the molecular weight of polymer chains\cite{Noda:1981fk}. $\xi$ being the correlation length of concentration fluctuations, it can be measured by SANS. Solutions of PEG in heavy water of different volume fractions, $\phi_{\text{PEG}}$, have been characterized in this way at $T=20^\circ\text{C}$. Spectra are plotted in Fig.\ref{figspectre}. They were fitted using a lorentzian shape following the Ornstein-Zernike approximation expected for linear polymer in semi-dilute solution\cite{deGennes:1996}:
\begin{equation}\label{eqoz}
I(q) = \frac{I(0)}{1+(q\xi)^2}
\end{equation}
where $I$ is the coherent differential scattering cross section per unit volume and $q$ the scattering vector. The mass of a chain segment of size $\xi$ is $m(\xi)= I(0)/K^2\phi_{\text{PEG}}$, where $K^2=4.3\times 10^{-3}$\,cm$^{-1}$g$^{-1}$mol is the contrast factor of PEG in heavy water. If $a$ and $m_a$ denote the size and the mass of the monomer, for a semi-dilute solution $\xi/a$ and $g=m(\xi)/m_a$  can be interpreted as the mean short distance and the curvilinear distance between neighboring polymer chains\cite{deGennes:1996}, both expressed in monomer unit. Results are plotted in Fig.\ref{figxi}. We find $\xi/a=(1.10\pm0.02)\phi_{\text{PEG}}^{-0.75\pm 0.03}$ and $g=(1.14\pm0.02)\phi_{\text{PEG}}^{-1.31\pm 0.01}$ in very good agreement with the scaling laws predicted for semi-dilute polymers in good solvent\cite{deGennes:1996}, but also with the measurement of second virial coefficients of PEG reported in the literature\cite{SCWang:2002}.

SANS measurements were performed in heavy water. For chains in athermal solvent, isotopic substitution has no effect on static properties of the solution that remain inherent in the chemical structure of chains. However, water is not athermal for PEG and the local conformation of chains certainly differs in $^2$H$_2$O and $^1$H$_2$O. In most cases, isotopic substitution amounts to shift transition temperatures. For PEG in water, a linear decrease of the osmotic pressure with temperature has been reported\cite{Stanley:2003fk} for semi-dilute solutions. Following Eq.\ref{pi}, the corresponding variation of $\xi$ with temperature can be evaluated as:
\begin{equation}\label{xideT}
\xi(\phi,T)=\xi(\phi,20^\circ\text{C})\times\left({\frac{\pi(\phi,T)}{\pi(\phi,20^\circ\text{C})}}\right)^{1/3}
\end{equation}
Results reported by Stanley \& Strey\cite{Stanley:2003fk} correspond to an increase of $\xi$ by upmost 25\% in our temperature range. Isotopic substitution is expected to have an even smaller effect.

\subsection{Circular dichroism}

\begin{figure}[!htbp]
\centering
\includegraphics[width=1\linewidth]{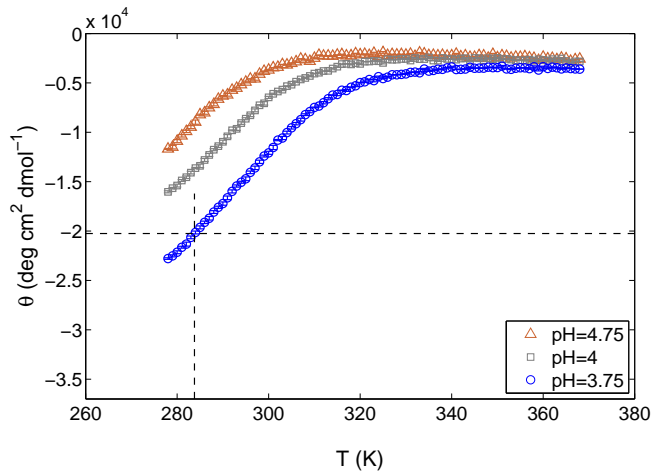}
\caption{Molar ellipticity $\theta$ at 222\,nm per amino acid of poly(L-glutamic acid) solutions vs. temperature for different pH. The horizontal dashed line has an ordinate $\theta_{\text{1/2}}$ corresponding to the helix fraction $1/2$.}
\label{figph}
\end{figure}

Circular dichroism measurements have been performed on poly(L-glutamic acid) aqueous solutions. We focused on the molar ellipticity per amino acid, $\theta$, measured at 222\,nm that allows us to deduce the fraction of amino acid of poly(L-glutamic acid) involved in $\alpha$-helices. In Fig.\ref{figph}, the ellipticity measured as a function of temperature for heating rate equal to 1\,deg/min is plotted for different pH values. In this figure, the horizontal dashed line indicates the value $\theta_{\text{1/2}}=\theta(T^*)=(\theta(0)-\theta(\infty))/2+\theta(\infty)$ corresponding to the fraction $1/2$ of helix amount at the transition temperature $T^*$. The asymptotic value $\theta(\infty)$ was estimated from our measurements as equal to $-3500$\,deg cm$^2$ dmol$^{-1}$, whereas $\theta(0)$ (corresponding to 100\% of helix) was taken as equal to $-3.7\times10^4$\,deg cm$^2$ dmol$^{-1}$ according to ref.\cite{Su:1994}. The intercept of each experimental curve with this line allows us to determine the transition temperature $T^*$. The lower the ellipticity, the higher the helix amount, thus our measurements indicate clearly that for poly(L-glutamic acid) at a given temperature, the helix amount strongly decreases with increasing pH. Actually by increasing pH, carboxylic acid groups are dissociated leading to electrostatic repulsions between amino acids and thus to an increase of the effective bond enthalpy $H_B$ of helix formation. This results in an increase of the transition temperature $T^*$\cite{Nakamura:1981}. In this work, we have taken advantage of this feature to finely tune $T^*$ in the accessible temperature-window for experiments at atmospheric pressure. In Fig.\ref{figph}, we can see that by decreasing pH, the transition begins to enter the accessible temperature-window at pH=3.75, allowing an eventual stabilization even more pronounced to be studied. Note that at even smaller pH, poly(L-glutamic acid) clearly aggregates and precipitates. In this regime, this experimental issue is clearly evidenced by heating and cooling curves that do not superimpose. It can be reasonably assumed that at pH=3.75, a possible aggregation should be also revealed by a similar behavior. At pH=3.75, with a temperature ramp of 1\,deg/min, the ellipticity displays an hysteresis smaller than 5\% that allows us to reasonably assume that the solution remains free of aggregates.

\begin{figure}[!htbp]
\centering
\includegraphics[width=1\linewidth]{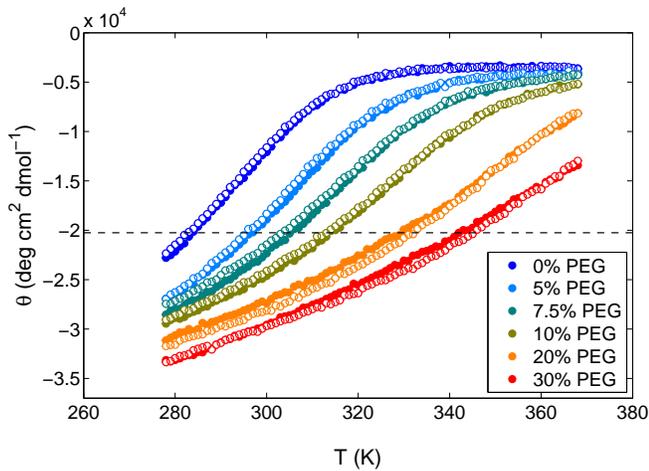}
\caption{Ellipticity $\theta$ at 222\,nm per amino acid of poly(L-glutamic acid) solutions at pH=3.75 vs. temperature for different volume fractions of PEG. Full and open symbols correspond to rising and decreasing temperature curves, respectively. The horizontal line has an ordinate $\theta_{\text{1/2}}$ corresponding to the helix fraction $x_h=1/2$.}
\label{figcd1}
\end{figure}

At pH=3.75, circular dichroism measurements have been performed as a function of temperature for different volume fractions of PEG. Results are plotted in Fig.\ref{figcd1}. Measurements indicate plainly the stabilization of helices with PEG addition. Despite that no hysteresis has been practically observed, the further data analysis were done only on rising-temperature curves in order to avoid possible smearing effects. 

\begin{figure}[!htbp]
\centering
\includegraphics[width=1\linewidth]{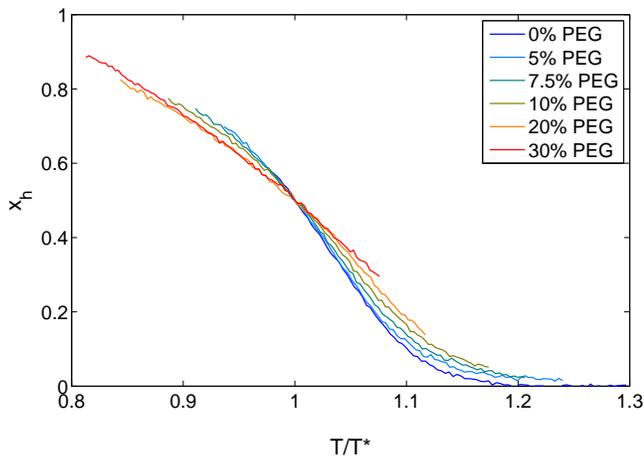}
\caption{Helix fraction $x_h$ of poly(L-glutamic acid) solutions at pH=3.75 vs. reduced temperature $T/T^*$ for different volume fractions of PEG.}
\label{figcd2}
\end{figure}

From the ellipticity measurements reported in Fig.\ref{figcd1}, the helix fraction $x_h$ was computed following the relation:
\begin{equation}\label{eqcd}
x_h(T)=\frac{\theta(T)-\theta(\infty)}{\theta(0)-\theta(\infty)}
\end{equation}
In Fig.\ref{figcd2}, $x_h$ is plotted as a function of the reduced temperature $T/T^*$, where $T^*$ is determined by the intercept of each experimental curve with the $\theta_{\text{1/2}}$ value. 

\section{Discussion}

Current theories on helix-coil transition only address the case of a single-molecule (volume fraction $\phi_{\text{peptide}}\to 0$ limit with no peptide-peptide interactions) of infinite length (number of amino acids $N_{\text{peptide}}\to\infty$). In the present study, we intended to work in this regime. Actually:

\begin{itemize}
\item Our measurement were performed in very dilute solution of peptide ($\phi_{\text{peptide}}=2\times 10^{-4}$). In our knowledge in this concentration range, the effect of peptide-peptide interactions on the helix-coil transition has never been observed. 

\item The effect of the peptide length on helix/coil transition has been studied\,\cite{Hansmann:1999uq} and shown to be relevant for much more shorter chains as in our study. In this paper $N_{\text{peptide}}=150$, which we assume to be enough to correspond to the $N_{\text{peptide}}\to\infty$ limit (note that the mean helix length at the transition that is reported below is much smaller than $N_{\text{peptide}}$ and is thus consistent with this assumption).
\end{itemize}

Our results will be discussed assuming a two-state transition between coil and helix. This point is not obvious given that three states (or more) could be introduced (see e.g. ref.\cite{Lifson:1961}). However, this paper is concerned with the effect of crowding on the transition rather than a detailed analysis of its mechanism. For our purpose the two-state hypothesis is sufficient. Note that in Fig.\ref{figcd2}, the curves obtained for different volume fractions of PEG do not superimpose. Without further analysis, this demonstrates that the helix-coil transition cannot be described by the Van't Hoff equation for chemical equilibria, given that the latter would imply $T/T^*$ as being a reduced variable.

\subsection{Zimm-Bragg model} The first and simplest model concerned with the helix-coil transition is due to B. H. Zimm and J. K. Bragg\cite{zimm:526}. This model is widely used to describe experimental data and a basis of many refinements. In terms of the Zimm-Bragg model, taking the coil state as reference, the Gibbs free energy of a given monomer $i$ writes:
\begin{equation}
G_i=-hs_i-Js_i s_{i+1}
\end{equation}
with $s_{i}=0$ or $1$ for coil and helix segments, respectively. Here, $h$ corresponds to the free energy cost of a helical nucleus obtained by the formation of one hydrogen bond between $i$ and $i+4$ monomers: 
\begin{equation}\label{eqng}
h=H_{\text{B}}-TS_4
\end{equation}
where $H_{\text{B}}$ is the bond enthalpy and $S_4$ the entropy loss resulting from fixing 4 monomers in a helix turn. For charged monomers, $H_B$ is an effective bond enthalpy that accounts also for their electrostatic repulsions. The coupling parameter $J$ accounts for a loss of entropy $S_1$ that is smaller in the case of adding only one monomer to a pre-existing helix turn. This term is purely entropic and can be written as:
\begin{equation}\label{eqnj}
J=T(S_4-S_1)
\end{equation}
It can be viewed as the free energy cost of the helix extremities. Traditionally, the Zimm-Bragg parameters $s$ and $\sigma$ are introduced in order to express the statistical weights:
\begin{align}
\sigma s & = e^{h/kT}\label{eqnsigmas} \\ 
s & = e^{(h+J)/kT} =e^{(H_{\text{B}}-TS_1)/kT}\label{eqns}
\end{align}
corresponding to helix nucleation and propagation, respectively. One gets for $\sigma$:
\begin{equation}
\sigma=e^{-J/kT} \label{eqnsig}
\end{equation}
Directly from the above definitions, one can see that $s$ is temperature dependent, whereas $\sigma$ is not. For long chains, as a function of these parameters, the helix fraction $x_h$ and the extent $N_h$ of helical domains (average number of amino acids per helix block) write\cite{Grosberg_Khokhlov_1994}:
\begin{align}
x_h & =\frac{1}{2}+\frac{s-1}{ 2\left({ (s-1)^2+4 \sigma s }\right)^{1/2} }\label{xh}\\
N_h & =1+\frac{2s}{ 1-s+\left({ (s-1)^2+4 \sigma s }\right)^{1/2} } \label{xih}
\end{align}
The fraction $x_h(s)$ displays a sigmoid shape. The slope $\left({dx_h/ds}\right)_{\frac{1}{2}}$ at the very point $(x_h=1/2, s=1)$ of equal helix and coil amounts defines the abruptness of the transition. Eq.\ref{xh} gives $\left({dx_h/ds}\right)_{\frac{1}{2}}=1/4 \sigma^{1/2}$. 

\subsection{Data analysis}
Experimentally, $x_h(T)$ should be more conveniently used. However due to the primitiveness of the Zimm-Bragg model, a direct data fitting using Eq.\ref{xh} expressed as a function of temperature leads usually to poor results (see for instance ref.\cite{baldwin:1995}). In the literature, many improvements of the Zimm-Bragg model are proposed (see for instance \cite{Lifson:1961, chen:2007, Vorov:2009vl, Murza:2009ys}). For our part, we focused our data analysis on the near vicinity of the transition temperature and assumed that, in this narrow temperature range, the Zimm-Bragg model captures the main features of the helix-coil transition. The transition temperature corresponding to $(x_h=1/2, s=1)$ writes:
\begin{equation}\label{eqnstar}
T^*=H_B/S_1
\end{equation}
Eq.\ref{eqns}, \ref{xh} and \ref{eqnstar} yield:
\begin{equation}\label{slope}
\left({-\frac{dx_h}{d(T/T^*)}}\right)_{T=T^*} = \frac{S_1}{k}\times\frac{1}{4\sigma^{1/2}}
\end{equation}
In practice, $S_1/k$ has been measured as being of the order of 1\cite{Shi:2002fk,Ohkubo:2003uq}. This value will be retained in the following.
Note that at the transition temperature $T^*$, the extent of helical domains remains finite and from Eq.\ref{xih} equal to:
\begin{equation}\label{xihtstar}
N_h(T^*)=1+\sigma^{-1/2}
\end{equation}
Helices being rod-like, their extent allows us to define a correlation length $L_h$ of helical domains:
\begin{equation}
L_h=\frac{1}{2}\times\frac{N_h}{3.6}\times 5.4\text{\AA}
\end{equation}
since the repeat unit of $\alpha$-helix corresponds to 3.6 residues and 5.4\,\AA. Similarly to the notion of persistence length vs. Kuhn length in polymer physics, the factor $1/2$ comes from correlations that are symmetrical with respect to an arbitrary origin. 

Summary of our results for $T^*$, $\sigma$, $N_h$ and $L_h$ are reported in Table \ref{table1}.
For increasing concentration of PEG, we observe a systematic variation of the slope of $x_h(T/T^*)$ at $T^*$ (see Fig.\ref{figcd2}) that corresponds to an increase of the Zimm-Bragg parameter $\sigma$ and a corresponding decrease of the correlation length $L_h$ of helical domains. Note that the model-dependent values here reported for $L_h$ are in very good agreement with the value directly measured by Muroga et al.\cite{Muroga:1988fk} at $\phi_{\text{PEG}}=0$ by small angle X-ray scattering on poly(L-glutamic acid) at the transition temperature.

\begin{table}[!htbp]
\caption{\label{table1}Summary of the results obtained by circular dichroism measurements: transition temperature $T^*$; Zimm-Bragg parameter $\sigma$ deduced from the slope at $T^*$ of $\theta(T)$; extent $N_h$ and correlation length $L_h$ of helical domains at the transition.}
\begin{ruledtabular}
\begin{tabular}{c|c|c|c|c}
 $\phi_{\text{PEG}}$ & $T^*$ & $\sigma$ & $N_h$ & $L_h$\\
 \% & K & $\times 10^3$ &  & \AA\\
\hline
0    & 284& $3.45\pm 0.05$ & $18.1\pm 0.1$ & $13.6\pm 0.1$\\
5 & 297& $4.1\pm 0.1$ & $16.6\pm 0.2$ & $12.4\pm 0.2$\\
7.5 & 305& $4.8\pm 0.1$ & $15.4\pm 0.2$ & $11.6\pm 0.2$\\
10 & 313& $6.4\pm 0.2$ & $13.5\pm 0.2$ & $10.4\pm 0.1$\\ 
20 & 330& $8.3\pm 0.3$ & $12.0\pm 0.2$ & $8.8\pm 0.1$\\
30 & 342& $8.8\pm 0.2$ & $11.6\pm 0.1$ & $8.7\pm 0.1$\\
\end{tabular}
\end{ruledtabular}
\end{table}

\subsection{Variation of the solvent quality} 
Let us first consider a possible change of the solvent quality due to the increase of PEG concentration, i.e. the co-solute. 
Such effect of the co-solute on the helix-coil transition has been theoretically studied in ref.\cite{Pincus:2002}. The authors classify the solvent quality changes into two categories. The first one acts on the Zimm-Bragg parameter $s$ resulting in a shift of the transition temperature (e.g. variation of the pH enters in this category). The second category concerns the variation of the hydrogen-bonding ability of the solvent due to the co-solute. For instance, one may imagine that the co-solute preferentially binds to helix extremities, lowering their free energy and then increasing the Zimm-Bragg parameter $\sigma$ but remaining $s$ unaffected. 
From ref.\cite{Pincus:2002}, one would expect:
\begin{equation}\label{pincus}
\sigma(\phi)^{1/2}=\sigma(0)^{1/2}+e^{-\tilde{J}/kT} \phi 
\end{equation}
where $\tilde{J}$ is the free energy of a helix extremity in contact with the co-solute with a probability $p_{\text{contact}}=\phi$. 
Of course, the co-solute could also bind to the helix body (but with a different affinity constant than for the extremities), resulting in a change of the parameter $s$ and a shift of the transition temperature, such as the co-solute could be responsible for effects of both categories. However, a change of the solvent quality with PEG addition should be independent of the chain length and in particular should be observed for small PEG chains. This is not the case. In ref.\cite{Stanley2008}, on the same poly(L-glutamic acid) system but with short PEG chains (400\,g/mol), the authors report a stabilization of the helices (as in our experiments), but no change of cooperativity of the transition was reported.

\subsection{Entropic stabilization} 
\begin{figure}[!htbp]
\centering
\includegraphics[width=1\linewidth]{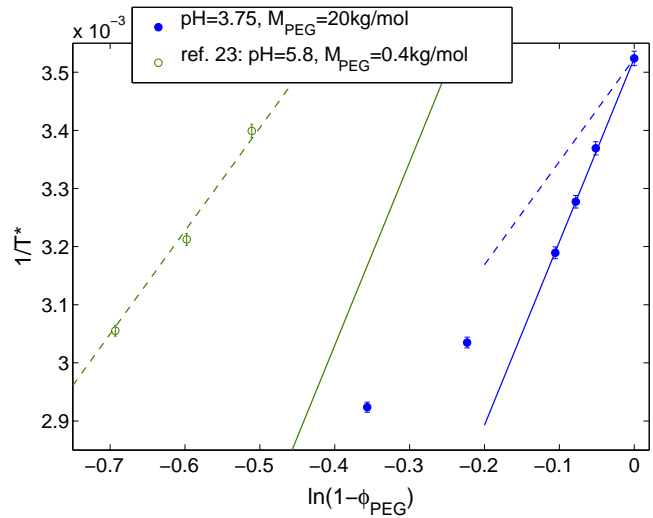}
\caption{Reverse transition temperature $1/T^*$ vs. logarithm of the accessible volume fraction $\ln(1-\phi_{\text{PEG}})$. Straight lines are guides for the eyes with slopes equal to 1/560\,K\cite{hermans:1966} for fully protonated (dashed lines) and 1/317\,K\cite{Rifkind:1964uq} for fully dissociated (full lines) glutamic acid monomers, respectively. Full symbols: present work (pH=3.75). Open symbols: from Fig. 5 in ref.\cite{Stanley2008} (pH=5.8). At this latter pH, $T^*(0)$ (intercept of dashed and full lines)  is not accessible. It has been arbitrary set to 235\,K in order the green dashed line fits the data points of ref. \cite{Stanley2008}.}
\label{figts}
\end{figure}

The number of possible states of a peptide chain is proportional to the accessible volume that is decreased by the crowding agent. For a given volume fraction $\phi_{\text{PEG}}$, the entropy of the peptide chain is:
\begin{equation}\label{eqnS}
S(\phi_{\text{PEG}})=S(0)+k\ln(1-\phi_{\text{PEG}})
\end{equation}
The basic idea to account for the impact of a crowding agent on structural transitions amounts to summarize its action to this "excluded volume effect" on the entropy.
In this framework, the equilibrium temperature $H_{\text{B}}/S(\phi_{\text{PEG}})$ is expected to vary as:
\begin{equation}\label{eqnT}
\frac{1}{T^*(\phi_{\text{PEG}})}=\frac{1}{T^*(0)}+\frac{k}{H_{\text{B}}}\ln(1-\phi_{\text{PEG}})
\end{equation}
In Fig.\ref{figts}, the reverse transition temperature $1/T^*$ is plotted as a function of $\ln(1-\phi_{\text{PEG}})$ for the data obtained at pH=3.75 (full symbols). In this figure, the straight lines are guides for the eyes with slopes taken from the literature as equal to $k/H_{\text{B}}=1/560$\,K (i.e. $H_{\text{B}}=1120$\,cal/mole from ref.\cite{hermans:1966} for fully protonated glutamic acid monomers) and $k/H_{\text{B}}=1/317$\,K (i.e. $H_{\text{B}}=630$\,cal/mole from ref.\cite{Rifkind:1964uq} for fully dissociated glutamic acid monomers), respectively. At least in the regime of low concentrations of PEG, our results seem in better agreement with this latter value rather than with the former. At pH=3.75, poly(L-glutamic acid) is certainly not fully, but only partly dissociated (the pKa of poly(L-glutamic acid) is of the order of 2.1). Thus, this result could indicate a co-solute effect of the first category that was discussed in the previous section, i.e. PEG probably lowers a bit the effective hydrogen bond enthalpy $H_{\text{B}}$.

In Fig.\ref{figts}, our data are compared to those measured by Stanley \& Sprey\cite{Stanley2008} at pH=5.8 (open symbols) for small PEG chains ($M_{\text{PEG}}=400$\,g/mol). At this pH, helices are much less stable and transition temperatures are shifted to lower values and cannot be directly compared to ours. However, their variations with $\phi_{\text{PEG}}$ should be similar and should also obey Eq.\ref{eqnT}. One can see that from this point of view, both sets of data are quite comparable and equally compatible with the values of $H_{\text{B}}$ reported in the literature.

Let us put the emphasis on this latter point. By finding $1/T^*$ vs. $\phi_{\text{PEG}}$ consistent with Eq.\ref{eqnT} and the value of $H_{\text{B}}$ reported in the literature, we show that the solvent quality with respect to poly(L-glutamic acid) does not vary significantly with PEG addition but also with temperature (at least between 280 and 340\,K, i.e. the range of measured $T^*$).

From Eq.\ref{eqnj}, the coupling parameter $J$ (and $\sigma$ defined by Eq.\ref{eqnsig}) results from a difference of two entropy terms that both are increased by the same quantity with PEG addition (Eq.\ref{eqnS}). Thus, an entropic stabilization mechanism alone would allow us to expect that $\sigma$ is independent of $\phi_{\text{PEG}}$, which is clearly in disagreement with our finding. Thus, a supplementary mechanism has to be proposed. 

\subsection{Polymeric nature of the crowding agent} 
\begin{figure}[!htbp]
\centering
\includegraphics[width=1\linewidth]{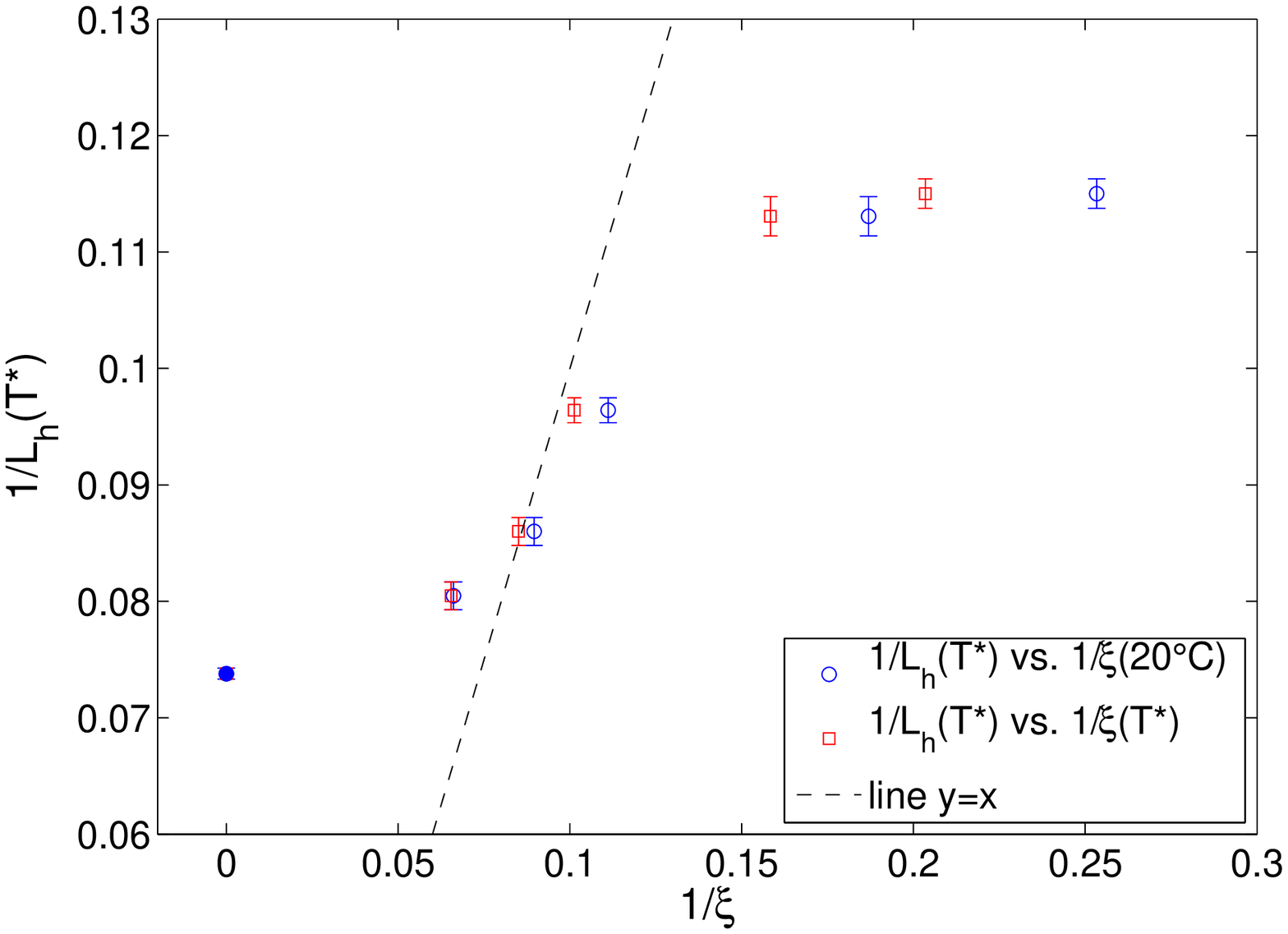}
\includegraphics[width=1\linewidth]{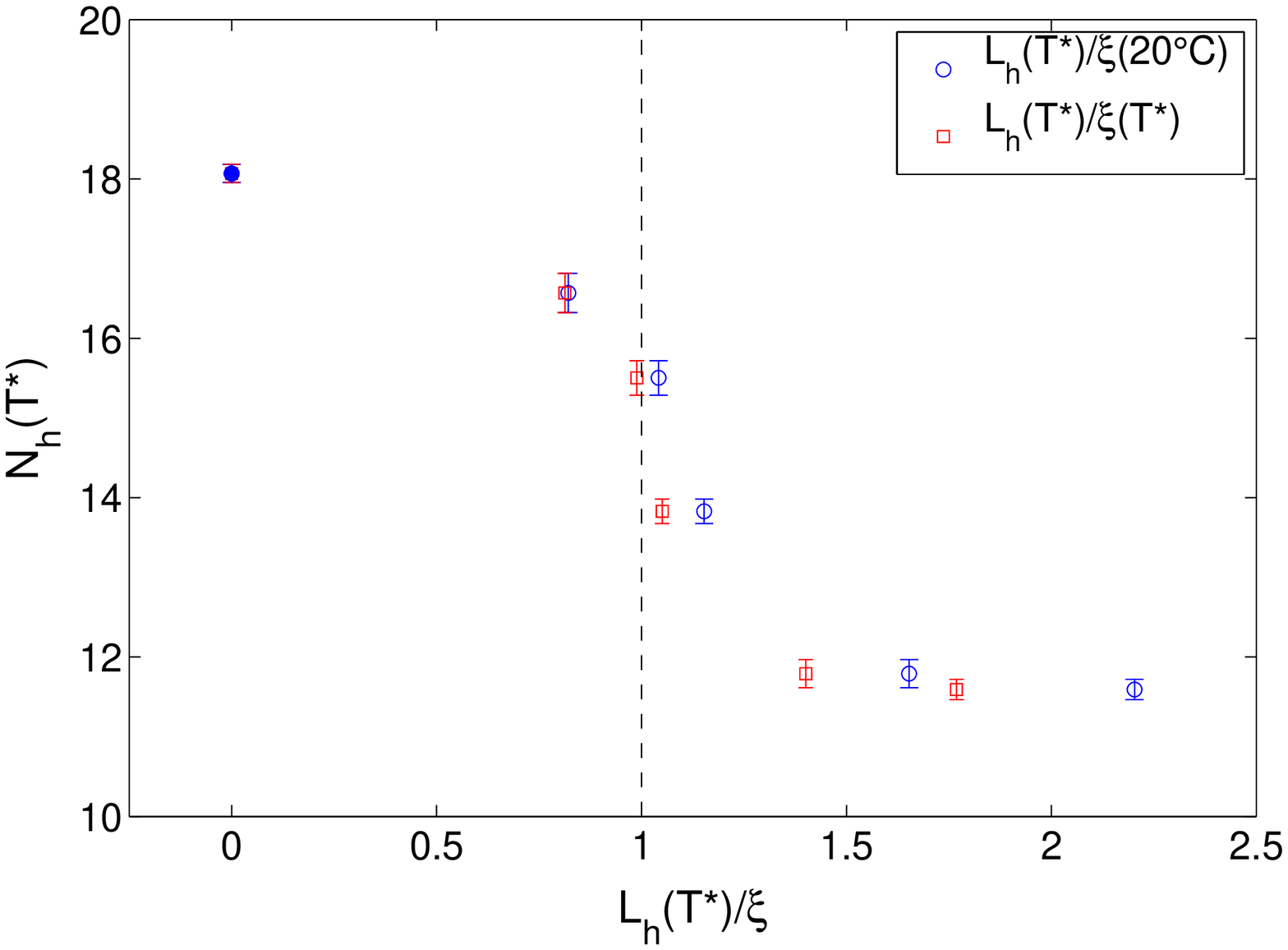}
\caption{Top: reverse correlation length $L_h(T^*)$ of helices at the transition vs. reverse mesh size $\xi$ of the PEG network. Bottom: average number $N_h(T^*)$ of amino acid per helix at the transition vs. the ratio $L_h(T^*)/\xi$. The close symbol at x-coordinate equal to 0 corresponds to $\phi_{\text{PEG}}=0$ (i.e. $\xi\to \infty$). Either $\xi$ measured at $20^\circ\text{C}$ (circles) or $\xi(T^*)$ (squares) calculated from Eq.\ref{xideT} and ref.\cite{Stanley:2003fk} are used to compute the x-coordinate.}
\label{Lh}
\end{figure}

In this paper, we would like to point out the polymeric nature of PEG as a crowding agent, a point that has not been considered so far. In this concentration range, the static properties (i.e. thermodynamics, structure) of PEG solutions depend on a single characteristic length, which so naturally can be compared to the correlation length of the helical domains. In Fig.\ref{Lh}, the average number $N_h(T^*)$ of amino acid per helix at the transition is plotted as a function of the ratio of the only two lengths $L_h(T^*)/\xi$ in our system. In this figure, either $\xi$ measured at $20^\circ\text{C}$ or $\xi(T^*)$ calculated from Eq.\ref{xideT} and ref.\cite{Stanley:2003fk} are used to compute the x-coordinate. The result strongly suggests two regimes ($L_h(T^*)\ll\xi$ and $\xi\ll L_h(T^*)$ separated by a wide crossover.

Long PEG chains are semi-dilute and form a network of characteristic length $\xi$ that can be viewed as a mean distance between two chains. Thus by definition, at length scales below $\xi$, a chain is alone and only the interactions between monomers belonging to the same chain are relevant. In contrast, beyond $\xi$, a given monomer experiences interactions and contacts with other chains. In good solvent, classically this results in a swollen conformation of the chain below $\xi$ (monomers of a given chain repeal each other) that becomes gaussian beyond $\xi$ (repulsions are screened due to the presence of other chains)\cite{deGennes:1996}. In the absence of any specific interaction between dilute chains of poly(L-glutamic acid) chains and semi-dilute PEG, the first are fully embedded and participate in the PEG network. Fig.\ref{Lh} may suggest that in the same way as local swelling of PEG chains is screened above $\xi$, the thermodynamics of helix growth could also be influenced by binary-contacts with the PEG network. 

Note that in this picture, for a given PEG concentration and for decreasing temperatures, growing helices begin to experience the network at a given temperature such as $L_h=\xi$. So that with PEG, $x_h=f(T)$ cannot be fitted in the entire temperature range with a model (such as the Zimm-Bragg model) that does not introduce the characteristic size of the network.

\section{Conclusion} 
We report circular dichroism measurements on the helix-coil transition of poly(L-glutamic acid) in semi-dilute solutions of PEG as a crowding agent that has been fully characterized by small angle neutron scattering. We show that the increase of PEG concentration stabilizes the helices and increases the transition temperature. This point, which has been already reported for other crowding agent species, is in agreement with an "entropic stabilization mechanism". However, we also notice that the increase of crowding agent concentration reduces the mean helix extent at the transition, or in other words reduces its cooperativity. To our knowledge, this result has not been reported previously and cannot be accounted for by an entropic stabilization.
Comparing the two lengths of the system, i.e. mean length of helices at the transition and mesh size of PEG network, our results strongly suggest two regimes: helices shorter or longer than the mesh size, each regime having its own value for the Zimm-Bragg parameter $\sigma$ that characterizes the cooperativity of the transition. 

\textit{A posteriori}, our results are not so surprising. Helix-coil transition is described by introducing a mean length of helical domains. Here, we have introduced a second length scale in the system with the PEG network. We observed that the growth of the former is hindered by the latter. In a sense, our result makes the physics of crowding-effects in some cases closer to that of phase transition in confined media. Until now, these kinds of effect have not been considered from a theoretical point of view and we hope that our measurements will contribute to stimulate works in this direction.


%

\end{document}